\documentclass[aps,prd,superscriptaddress,showpacs,showkeys,nofootinbib,floatfix]{revtex4}
\usepackage{graphicx}
\usepackage{bm}
\usepackage{color}
\usepackage{url}
\usepackage{amsmath}
\usepackage{amssymb}
\usepackage{mathrsfs}
\usepackage[lofdepth,lotdepth,caption=false]{subfig}
\usepackage{bm}
\usepackage{accents}
\usepackage{mathrsfs}

\newcommand{\eq}[1]{\begin{eqnarray}#1\end{eqnarray}}

\begin{document}

\title{Constraining the Violation of Equivalence Principle with IceCube Atmospheric Neutrino Data} 
\author{A. Esmaili}
\email{aesmaili@ifi.unicamp.br}
\affiliation{Instituto de F\'isica Gleb Wataghin - UNICAMP, 13083-859, Campinas, SP, Brazil\ }
\affiliation{Institute of Convergence Fundamental Studies, Seoul National University of Science and Technology, Gongreung-ro 232, Nowon-gu, Seoul 139-743, Korea}
\author{D. R. Gratieri}
\email{gratieri@ifi.unicamp.br}
\affiliation{Instituto de F\'isica Gleb Wataghin - UNICAMP, 13083-859, Campinas, SP, Brazil\ }
\affiliation{High and Medium Energy Group, Instituto de F\'isica e Matem\'atica, Universidade Federal de Pelotas, Caixa Postal 354, CEP 96010-900, Pelotas, RS, Brazil}
\author{M. M. Guzzo}
\email{guzzo@ifi.unicamp.br}
\affiliation{Instituto de F\'isica Gleb Wataghin - UNICAMP, 13083-859, Campinas, SP, Brazil\ }
\author{P. C. de Holanda}
\email{holanda@ifi.unicamp.br}
\affiliation{Instituto de F\'isica Gleb Wataghin - UNICAMP, 13083-859, Campinas, SP, Brazil\ }
\author{O. L. G. Peres}
\email{orlando@ifi.unicamp.br}
\affiliation{Instituto de F\'isica Gleb Wataghin - UNICAMP, 13083-859, Campinas, SP, Brazil\ }
\affiliation{Abdus Salam International Centre for Theoretical Physics, ICTP, I-34010, Trieste, Italy\ }
\author{G. A. Valdiviesso}
\email{gustavo.valdiviesso@unifal-mg.edu.br}
\affiliation{Instituto de Ci\^encia e Tecnologia, Universidade Federal de Alfenas, Unifal-MG, Rod. Jos\'e Aur\'elio Vilela, 11999, 37715-400 Po\c{c}os de Caldas MG, Brazil}

\date{\today} 

\begin{abstract}
The recent high-statistics high-energy atmospheric neutrino data collected by IceCube open a new window to probe new physics scenarios that are suppressed in lower energy neutrino experiments. In this paper we analyze the IceCube atmospheric neutrino data to constrain the Violation of Equivalence Principle (VEP) in the framework of three neutrinos with non-universal gravitational couplings. In this scenario the effect of VEP on neutrino oscillation probabilities can be parametrized by two parameters $\Delta \gamma_{21}\equiv \gamma_2-\gamma_1$ and $\Delta\gamma_{31}\equiv \gamma_3-\gamma_1$, where $\gamma_i$'s denote the coupling of neutrino mass eigenstates to gravitational field. By analyzing the latest muon-tracks data sets of IceCube-40 and IceCube-79, besides providing the 2D allowed regions in $(\phi\Delta\gamma_{21},\phi\Delta\gamma_{31})$ plane, we obtain the upper limits $|\phi\Delta\gamma_{21}| < 9.1\times 10^{-27}$ (at 90\% C.L.) which improves the previous limit by $\sim4$ orders of magnitude and $|\phi\Delta\gamma_{31}| \lesssim 6\times 10^{-27}$ (at 90\% C.L.) which improves the current limit by $\sim1$ order of magnitude. Also we discuss in detail and analytically the effect of VEP on neutrino oscillation probabilities.

\end{abstract}

\maketitle

\section{Introduction}\label{introduction}

The Equivalence Principle is the cornerstone of classical gravitational theories, from Newtonian gravitation to General Relativity. The Weak Equivalence Principle (WEP) states that the geodesic paths followed by free falling bodies are the same, regardless of their energy content. In the other words, the motion of a falling body is determined only by the surrounding geometry and not by the body's own properties~\cite{Gravitation}. In the weak field limit, this principle leads to an universal acceleration of the falling bodies, a fact that is rooted in two principles of the Newtonian gravitation:  the equivalence of inertial and gravitational masses and universality of the Newton's gravitational constant $G_N$. Since the proposal of WEP, this hypothesis has been extensively tested by a large diversity of experiments, including torsion-balance experiments~\cite{Eot-wash}, motion of solar system bodies~\cite{Overduin:2013soa}, spectroscopy of atomic levels~\cite{Hohensee:2013cya} and pulsars~\cite{Damour:1991rq,Horvat:1998st,Barkovich:2001rp}; which always lead to strong limits on possible deviations. However, recent developments in theoretical physics are systematically indicating that many modern attempts to obtain a quantum version of the gravitational theory lead to the prediction that the equivalence principle will be violated in some scale (see for example~\cite{Damour:2001fn,Damour:2002mi,ArmendarizPicon:2011ys,Damour:2010rm,Carroll:2008ub,Olmo:2006zu,Adunas:2000zm}). In this sense, improving the current limits on the VEP provides a diagnostic tool in probing very high energy theories of quantum gravity, which are almost inaccessible to conventional experiments.

One of the methods to probe VEP is through the neutrino oscillation phenomena. The effect of VEP on neutrino oscillation was first studied by Gasperini~\cite{Gasperini:1988zf} and later developed by others in~\cite{Gasperini:1989rt,Halprin:1991gs}. The original model was intended to solve the solar neutrino problem~\cite{Pantaleone:1992ha,Butler:1993wi,Bahcall:1994zw,Halprin:1995vg,Mureika:1996ud,Mansour:1998nb,Gago:1999hi,Casini:1999kt,Majumdar:2000sd}, which is now in excellent agreement with the framework of massive neutrinos with the MSW effect~\cite{Mikheev:1986wj,Mikheev:1986gs}. However, despite its failure to explain the solar neutrino problem, VEP can contribute to flavor oscillation as a subdominant effect and so can be probed by solar neutrinos~\cite{Minakata:1994kt,Valdiviesso:2011zz}, atmospheric~\cite{Foot:1997kk,Foot:1998pv,Foot:1998vr,Fogli:1999fs,GonzalezGarcia:2004wg,GonzalezGarcia:2005xw,MACRO-ref,Morgan:2007dj,Abbasi:2009nfa}, supernova~\cite{Pakvasa:1988gd,Guzzo:2001vn}, cosmic~\cite{Minakata:1996nd} and accelerator~\cite{Iida:1992vh,Mann:1995nw} neutrinos.

Essentially, the sensitivity of neutrino oscillation to VEP originates from the fact that the flavor states of neutrinos are coherent superposition of mass eigenstates and so act as interferometers which are sensitive to differences in the coupling of mass states to gravitational field. The bottom line is that VEP effectively changes the mass-squared differences by adding a term proportional to the square of neutrino energy ($\propto E_\nu^2$). Thus, by the increase of neutrino energy the VEP effects become stronger and so the potential to discover/constrain VEP increases. Among the known perpetual sources of neutrinos, atmospheric neutrino energies extends up to very high energy and so provides a unique opportunity to probe VEP. The construction of huge (km$^3$ scale) neutrino telescopes, with the completed IceCube detector at the South Pole as an example, fulfills the detection of these high energy atmospheric neutrinos. Currently two sets of high energy atmospheric neutrino data are available from IceCube experiment: the ``IC-40" data set in the energy range (100 GeV - 400 TeV)~\cite{Abbasi:2010ie} and ``IC-79" data set in the range (20 GeV - 10 TeV)~\cite{Aartsen:2013jza}, with the total number of events: $\sim18,000$ and $\sim40,000$ respectively. In this paper we utilize these data in the search of VEP in the most general phenomenological model accommodating it. By analyzing these data we obtain the most stringent upper limit of VEP parameters, some of them are $\sim4$ orders of magnitude stronger than the current limits.   

This paper is organized in the following way: in section~\ref{model} we review the phenomenology of the oscillation of massive neutrinos in the presence of VEP and current upper limits on VEP parameters. In section~\ref{sec:vep-energy} we study in detail the effect of VEP on neutrino oscillation. Also, we show the numerical calculation of oscillation probabilities and their interpretation in terms of analytical approximations. Our analysis of the atmospheric neutrino data of IceCube is presented in section~\ref{analysis}. Conclusion is provided in section~\ref{sec:con}.

\section{Phenomenology of Massive Neutrinos in the Presence of VEP}\label{model}

Different approaches for the implementation of VEP in the neutrino sector of standard model exist. Originally, VEP was introduced as a mechanism which induces flavor oscillation even for massless neutrinos. For massless neutrinos, although the neutrino states do not couple directly to gravitational field, during the propagation the gravitational redshift develops a phase difference between the components of the superposition of gravitational eigenstates which leads to flavor oscillation~\cite{Gasperini:1988zf}. In this case the gravitational and flavor eigenstates do not coincide and are related to each other by a unitary mixing matrix. However, as it is confirmed by the data of more than two decades of neutrino oscillation experiments, neutrinos are massive with at least two different nonzero masses for the three mass eigenstates. Global analysis of oscillation data strongly verified that these mass differences are responsible for the observed oscillation phenomena and VEP (if exists) can contribute only sub-dominantly. Within this framework of massive neutrinos, three sets of eigenstates can be defined: mass eigenstates (which are defined by the diagonalization of charged lepton mass matrix), gravitational eigenstates (which diagonalizes the coupling matrix of neutrinos to gravitational field, the diagonalizing matrix is not proportional to unit matrix in the presence of VEP) and flavor eigenstates (which enter the charged current interaction). In general these three sets of eigenstates are not equal and so choosing one of them to write the Schr\"{o}dinger-like equation of evolution, demands to introduce two $3\times 3$ mixing matrix where one of them is almost the conventional PMNS matrix and the other parametrize the VEP (see~\cite{Gasperini:1989rt,Halprin:1995vg}). In this approach, the number of VEP parameters is equal to the number of parameters required to parametrize the $3\times3$ unitary matrix which is five, including the possible phases. However, since probing this multi-dimensional parameter space is cumbersome, we adopt a different approach which reduces the number of VEP parameters to two. The approach we adopt in this paper (which was introduced first in~\cite{Halprin:1991gs}) is based on the assumption that the weak equivalence principle is violated via the dependence of Newton's constant on the mass of the neutrino state; {\it i.e.}, $G_N'=\gamma_i\ G_N$, where $\gamma_i$ depends on the mass $m_i$ (so, $\gamma_i\rightarrow 1$ means restoration of equivalence principle). So, in our approach, VEP is induced by the non-universality of gravitational coupling among the neutrino states, which is effectively taken into account by modifying the metric in the weak field approximation. It is worth mentioning that since currently strong limits exist on VEP and no self-consistent quantum theory of gravity is envisaged, adopting this minimalistic and phenomenological approach is quite justifiable and robust.

In the weak field approximation, the space-time metric can be expressed as $g_{\mu\nu}=\eta_{\mu\nu}+h_{\mu\nu}(x)$ where the Minkowski metric $\eta_{\mu\nu}={\rm diag}(1,-1,-1,-1)$ and $h_{\mu\nu}=-2\gamma_i\phi(x)\delta_{\mu\nu}$~\cite{Will}. Here $\phi$ is the Newtonian gravitational potential and the constant $G_N$ is implicit. As we mentioned, VEP will be accommodated by introducing the multiplicative species-dependent factor $\gamma_i$ such that $\phi_{\rm VEP}=\gamma_i \phi$. Incorporating this metric in the Klein-Gordon equation (and thus neglecting the spin-flip effects), we obtain the following relation for the Hamiltonian eigenvalues~\cite{Valdiviesso:2011zz}:
\eq{
E_i & = &  p_\nu\ (1+2\gamma_i\phi) +\dfrac{m_i^2}{2p_\nu}(1+4\gamma_i\phi)~, 
\label{E_Halprin}
}
where $p_\nu$ denotes the neutrino momentum. The usual relativistic dispersion relation $E_i=p_\nu+m_i^2/(2p_\nu)$ is recovered for $\phi\ll 1$. Changing to the flavor basis, the Schr\"{o}dinger-like equation of neutrino evolution takes the following form:
\eq{
i\dfrac{{\rm d}\nu_\alpha}{{\rm d}r} & = & \left[ \dfrac{1}{2p_\nu}U\left( M^2 +\Delta G\right)U^\dagger +V(r)\right]_{\alpha\beta}\nu_\beta~,
\label{Sch}
}
where $U=U_{23}U_{13}U_{12}$ is the PMNS mixing matrix ($U_{ij}$'s are rotation matrices with angle $\theta_{ij}$, $i<j\leq3$) and the mass matrix $M^{2}={\rm diag}(0,\Delta m_{21}^2,\Delta m_{31}^2)$, where $\Delta m_{ij}^2\equiv m_i^2 - m_j^2$. In this equation $V(r)=\sqrt{2}G_F N_e(r){\rm diag}(1,0,0)$ is the effective matter potential of the Earth, where $G_F$ is the Fermi's constant and $N_e(r)$ is the electron number density profile of the Earth. Finally, the $\Delta G$ in Eq.~(\ref{Sch}) contains all the VEP contributions to neutrino oscillation and is given by
\eq{
\Delta G  = {\rm diag}\left(0,\pm 4p_\nu^2 \left| \phi(r) \Delta\gamma_{21}\right|,\pm 4p_\nu^2 \left| \phi(r) \Delta\gamma_{31}\right| \right)~,
\label{deltaG}
}
where the two VEP parameters $\Delta \gamma_{21}\equiv\gamma_2-\gamma_1$ and $\Delta\gamma_{31}\equiv\gamma_3-\gamma_1$ represent the differences between $G_N$ for the respective mass eigenstates. As can be seen, the observable VEP parameters are $\phi \Delta \gamma_{21}$ and $\phi\Delta \gamma_{31}$. With our current knowledge of the large scale structure of Universe, the dominant contribution to $\phi$ is from the Great Attractor with the value $\sim10^{-5}$; though ambiguities exist on this value and also on other possible sources. However, since the VEP effect appears just as the multiplication of $\phi$ and $\Delta\gamma_{ij}$, these ambiguities can be avoided by reporting the limits on $\phi \Delta \gamma_{ij}$ instead of $\Delta\gamma_{ij}$. Also, dominant contribution from large scale distant sources means that we can safely ignore the position dependence of $\phi(r)$ over the propagation path of atmospheric neutrinos and assume that the potential is constant. In Eq.~(\ref{deltaG}) the $\pm$ signs take into account the different possible hierarchies for VEP parameters $\gamma_i$; such that the plus sign means the hierarchy of VEP parameters is the same as the one exhibited by the masses, while the minus sign represents the case where the hierarchies do not match. Since there is no reason {\it a priori} to restrict these possibilities, we consider both the plus and minus signs in our analysis. The evolution equation of anti-neutrinos can be obtained from Eq.~(\ref{Sch}) by replacing: $V\to -V$ and $U\to U^\ast$.

In Table~\ref{ultima} we list the existing upper limits on VEP parameters $\phi\Delta \gamma_{ij}$ from various sources of neutrinos\footnote{It should be noticed that some of the limits on $\phi\Delta \gamma_{ij}$ in Table~\ref{ultima} have been obtained with the assumption that mass eigenstates and gravitational eigenstates are not equal and are related by a unitary transformation which in the $2\nu$ system can be parametrized by a rotation angle $\theta_{G}\in [0,\pi]$. The reported limits are either for $\theta_G=0$ or marginalized over $\theta_G$.}. As can be seen from this table the current upper limits on VEP parameters are $\phi\Delta \gamma_{32} \lesssim {\rm few} \times10^{-26}$ and $\phi\Delta \gamma_{21} \lesssim 10^{-22}$. The sensitivity of ANTARES and IceCube experiments have been studied in Refs.~\cite{Morgan:2007dj} and \cite{GonzalezGarcia:2005xw} respectively, with the result $\phi\Delta \gamma_{32} \lesssim 3\times 10^{-24}$ for ANTARES and $\lesssim 2\times 10^{-28}$ for IceCube. In this paper we derive the upper limits on $\phi\Delta \gamma_{ij}$ by analyzing the collected data by IceCube experiment.

\begin{table}[hhhh]
\caption{\label{ultima}Current upper limits on VEP parameters $\phi\Delta\gamma_{ij}$ from different analyses.}
\centering
\vspace{.5cm}
\begin{tabular}{|c|c|c|c|}
\hline
Neutrino source & $\phi\Delta\gamma_{32} $ & $\phi\Delta\gamma_{21}$ & Reference \\
\hline
\hline
SN1987A  & 0 & $ \lesssim 10^{-22}$ & \cite{Guzzo:2001vn} \\
Atmospheric (SK) & $ \lesssim 4\times 10^{-25}$ & 0 & \cite{GonzalezGarcia:2004wg}  \\
Atmospheric+K2K & $ \lesssim 6\times 10^{-26}$ & 0 & \cite{GonzalezGarcia:2004wg} \\
Atmospheric (MACRO) & $ \lesssim 3 \times 10^{-24}$ & 0 & \cite{MACRO-ref}\\ 
Atmospheric (AMANDA) & $ \lesssim 3\times 10^{-25}$ & 0 & \cite{Abbasi:2009nfa}\\
Solar & 0 & $ \lesssim 10^{-19}$ & \cite{Valdiviesso:2011zz}\\
\hline
\end{tabular}
\end{table}

Probabilities of flavor oscillation for atmospheric neutrinos propagating through the Earth can be obtained by the numerical solution of Eq.~(\ref{Sch}), with the matter density taken from PREM model of Earth~\cite{Dziewonski1981297}. For our analysis in section~\ref{analysis} we calculated these probabilities by scanning the parameter space of $\phi\Delta\gamma_{21}$ and $\phi\Delta\gamma_{31}$ and confront it with the published IC-40 and IC-79 data sets from IceCube neutrino telescope. However, before describing the analysis method, in the next section we discuss the signature of VEP in oscillation probabilities, especially in the high energy range ($E_\nu \gtrsim 100$~GeV) where IceCube collects data.

\section{Oscillation probabilities in the presence of VEP\label{sec:vep-energy}}

As can be seen from Eq.~(\ref{Sch}), VEP effectively modifies the standard neutrino oscillation picture by adding the term $\Delta G$ to the mass-squared matrix of neutrinos. Thus, basically the VEP in neutrino oscillation is equivalent to replacing the standard mass-squared differences $\Delta m_{ij}^2$ by
\begin{equation}
\Delta m^{2, {\rm eff}}_{ij}=\Delta m^{2}_{ij}\pm 4E^{2}_{\nu}\left| \phi\Delta\gamma_{ij}\right|~.
\label{eq:Dm2eff}
\end{equation}
Substituting this effective mass-squared difference in the evolution equation in Eq.~(\ref{Sch}), the first term in Eq.~(\ref{eq:Dm2eff}) which induces the standard oscillation is inversely proportional to $E_\nu$. It is well-known that the oscillation induced by this term diminishes in the high energy range ($\gtrsim 100$~GeV), which is our interest in this paper: the $\nu_e$-mixing is suppressed in high energy due to the Earth's matter effect, while the $\nu_{\mu/\tau}$ oscillation length $4\pi E_\nu/\Delta m_{31}^2 \sim10^5~{\rm km} \left( E_\nu/100~{\rm GeV} \right)$ becomes larger than the diameter of Earth $2R_\oplus\sim 12,000$~km and so $\nu_\mu-\nu_\tau$ oscillation will be suppressed. However, the second term in Eq.~(\ref{eq:Dm2eff}) which characterizes the VEP contribution to neutrino oscillation appears in the evolution equation as $2E_\nu\phi  \Delta\gamma_{ij}$ and so the effect of VEP dominates with the increase of energy. This dominant contribution of VEP to neutrino oscillation in the high energy range is the reason that neutrino telescopes, such as IceCube, are perfect detectors in probing VEP.

In the absence of VEP, since the matter effect suppresses $\nu_e$-oscillation, the oscillation of $\nu_{\mu/\tau}$ can be described by $2\nu$ approximation. In this approximation we can write the survival probability of muon neutrinos for the standard oscillation scenario as\footnote{Since we are interested in the high energy range ($\gtrsim 10$~GeV), here we neglect the parametric resonance and effect of $\theta_{13}$.}
\begin{equation}
P^{\rm std}(\nu_{\mu}\to \nu_{\mu})=1-\sin^{2}2 \theta_{23}\sin^{2}\left(\dfrac{\Delta m^{2}_{31}}{4E_{\nu}}L    \right)~,
\label{Pstd}
\end{equation}
where $L=-2R_\oplus\cos\theta_z$, with $R_\oplus$ and $\theta_z$ denoting respectively the Earth's radius and the zenith angle of incoming neutrino (for up-going neutrinos with $-1\leq\cos\theta_z\leq0$). The minima of the $\nu_\mu$ survival probability are at energies $E_{\nu,{\rm std}}^{{\rm min},n}$ where derive from the following condition ($n=0,1,\ldots$)
\begin{equation}
\frac{\Delta m^{2}_{31}L}{4E^{{\rm min},n}_{\nu,{\rm std}}}=\left(n+\frac{1}{2}\right) \pi \longrightarrow
E^{{\rm min},n}_{\nu,{\rm std}}=  24.8 \, {\rm GeV} \, \left(\frac{1}{2n+1}\right)\left(\frac{\Delta m^{2}_{31}}{2.41\times 10^{-3}\,{\rm eV}^2}\right)\left(\frac{\cos\theta_z}{-1}\right)~,
\label{emin}
\end{equation}
and the maxima are at $E_{\nu,{\rm std}}^{{\rm max},k}$ given by ($k=1,2,\ldots$) 
\begin{equation}
\frac{\Delta m^{2}_{31}L}{4E^{{\rm max},k}_{\nu,{\rm std}}} = k \pi \longrightarrow E^{{\rm max},k}_{\nu,{\rm std}} =  12.4\, {\rm GeV} \left(\dfrac{1}{k}\right)\left(\frac{\Delta m^{2}_{31}}{2.41\times 10^{-3}\,{\rm eV}^2}\right)\left(\frac{\cos\theta_z}{-1}\right)~.
\label{emax}
\end{equation}
Although the matter effect modifies this pattern, the first few maxima and minima can be read from Eqs.~(\ref{emin}) and (\ref{emax}) fairly. 

In the next subsections we extend this discussion to the case of VEP. We consider three cases: \textbf{Case \textit{i}}: $\Delta \gamma_{21}=0$ and $\Delta \gamma_{31}\neq0$; \textbf{Case \textit{ii}}: $\Delta \gamma_{21}\neq0$ and $\Delta \gamma_{31}=0$; and \textbf{Case \textit{iii}}: $\Delta \gamma_{21}=\Delta \gamma_{31}\neq0$.

\subsection{Case \textbf{\textit{i}} : $\phi\Delta\gamma_{21}=0$ and $\phi\Delta\gamma_{31}\neq0$}
\label{sub:casei}

Starting with the case where $\phi \Delta\gamma_{21}=0$, the VEP will modify only the $\Delta m^{2}_{31}$ as described by Eq.~(\ref{eq:Dm2eff}) such that\footnote{Since $\Delta m_{21}^2L/(4E_\nu)\ll1$ in the high energy range, we neglect this term and set $\Delta m_{21}^2=0$.}
\begin{equation}
\Delta m^{2,{\rm eff}}_{31}=\Delta m^{2}_{31}+4E^{2}_{\nu}\phi\Delta\gamma_{31}~~~~,~~~~\Delta m^{2,{\rm eff}}_{21}=0~.
\label{Dm2effi}
\end{equation}
In the analytical discussions of this section we assume $\Delta\gamma_{31}>0$, unless mentioned otherwise. Generalization to $\Delta\gamma_{31}<0$ is straightforward. In the propagation basis defined by $\left| \nu^{\prime}\right\rangle= U^{\dagger}_{23}\left| \nu \right\rangle$ evolution equation can be written as (since $\Delta m^{2,{\rm eff}}_{21}=0$, the $\theta_{12}$ angle can be neglected) 
\begin{eqnarray}
i\dfrac{\rm d}{{\rm d}t}
\left(
\begin{array}{cc}
\nu_{e}^{\prime}\\
\\
\nu_{\mu}^{\prime}\\
\\
\nu_{\tau}^{\prime}\\ 
\end{array}
\right)
=\left(
\begin{array}{ccc}
s^{2}_{13}\left(\dfrac{\Delta m_{31}^{2,{\rm eff}}}{2E_\nu}\right) + V & 
 0 & s_{13}c_{13} \left(\dfrac{\Delta m_{31}^{2,{\rm eff}}}{2E_\nu}\right)
\\
0 & 0 & 0\\ 
s_{13}c_{13} \left(\dfrac{\Delta m_{31}^{2,{\rm eff}}}{2E_\nu}\right)
&0 &c^{2}_{13}\left(\dfrac{\Delta m_{31}^{2,{\rm eff}}}{2E_\nu}\right) 
\\ 
\end{array}
\right)
\left(
\begin{array}{cc}
\nu_{e}^{\prime} \\
\\
\nu_{\mu}^{\prime}\\
\\
\nu_{\tau}^{\prime}\\ 
\end{array}
\right)~,
\label{rotate0}
\end{eqnarray}
where $c_{ij}=\cos \theta_{ij}$ and $s_{ij}=\sin \theta_{ij}$. In the above evolution equation $\nu_{\mu}^{\prime}$ decouples from the rest of states. For constant density, the flavor states $(\nu_{e}^{\prime}, \nu_{\mu}^{\prime},\nu_{\tau}^{\prime})$ at distance $L$ can be written as
\begin{align}
\left(
\begin{array}{cc}
\nu_{e}^{\prime} \\
\nu_{\mu}^{\prime}\\
\nu_{\tau}^{\prime}\\ 
\end{array}
\right)_{t=L}
=\left(
\begin{array}{ccc}
\mathcal{T}_{ee}       & 0 & \mathcal{T}_{e\tau}      \\
0             & \mathcal{T}_{\mu\mu} & 0\\
\mathcal{T}_{\tau e} & 0 & \mathcal{T}_{\tau\tau} \\ 
\end{array}
\right)
\left(
\begin{array}{cc}
\nu_{e}^{\prime} \\
\nu_{\mu}^{\prime}\\
\nu_{\tau}^{\prime}\\ 
\end{array}
\right)_{t=0}~,
\label{smatrix-casei}
\end{align}
with the $\mathcal{T}_{\alpha\beta}$ given by
\begin{equation}
\mathcal{T}_{ee,\tau\tau} =\cos\left( \dfrac{\Delta\widetilde{m}^2_{31}}{4E_\nu}L\right)\mp i \cos 2\widetilde{\theta}_{13}\sin\left(\dfrac{\Delta\widetilde{m}^2_{31}}{4E_\nu}L\right)  \quad , \quad \mathcal{T}_{e\tau}=\mathcal{T}_{\tau e}=-i \sin 2\widetilde{\theta}_{13}\sin\left(\dfrac{\Delta\widetilde{m}^2_{31}}{4E_\nu}L\right)~,
\label{Dmeff21-1}
\end{equation}
\begin{equation}
\mathcal{T}_{\mu\mu}= \exp \left[ i \dfrac{\left( \Delta m_{31}^{2,{\rm eff}}+2E_\nu V\right)}{4E_\nu} L\right]~,
\end{equation}
where
\begin{equation}
\Delta  \widetilde{m}^{2}_{31}=\sqrt{\left(\cos2\theta_{13}\Delta m_{31}^{2,{\rm eff}}
-2E_\nu V\right)^2+\left(\sin2\theta_{13}\Delta m_{31}^{2,{\rm eff}}   \right)^2}\quad , \quad
\sin 2\widetilde{\theta}_{13}=\sin 2 \theta_{13}
\dfrac{\Delta m_{31}^{2,{\rm eff}}}{\Delta\widetilde{m}^2_{31}}~.
\label{c2tms2tm-casei}
\end{equation}
Obviously a resonance can be identified in Eq.~(\ref{c2tms2tm-casei}) when $\cos2\theta_{13}\Delta m_{31}^{2,{\rm eff}}=2E_\nu V$, which occurs in neutrino (antineutrino) channel for $\phi\Delta\gamma_{31}>0$ ($\phi\Delta\gamma_{31}<0$). The resonance energy is
\begin{equation}\label{eq:res21}
E_\nu^{{\rm res},31}=\dfrac{V}{2\phi\Delta\gamma_{31}\cos 2\theta_{12}}\simeq 18~{\rm TeV} \left(\dfrac{10^{-26}}{\phi\Delta\gamma_{31}}\right) \left(\dfrac{0.9}{\cos2\theta_{13}}\right) \left(\dfrac{\langle \rho Y_e \rangle}{4.5\,{\rm gcm}^{-3}}\right)~,
\end{equation}
where $\langle\rho Y_e\rangle$ is the average density of Earth. The resonance is induced by $\theta_{13}$ angle and so it is absent when $\theta_{13}=0$. In the resonance region the $\widetilde{\theta}_{13}$ is maximal ($\simeq\pi/4$) and $\Delta\widetilde{m}_{31}^2$ has the minimum value ($\simeq\sin2\theta_{13}\Delta m_{31}^{2,{\rm eff}}$).

Rotating back to $\nu_\alpha$ flavor states, the $\nu_\mu$ survival probability from Eq.~(\ref{smatrix-casei}) is
\begin{align}
P (\nu_{\mu}\to\nu_{\mu}) &= \left| \left(U_{23} \mathcal{T} U_{23}^\dagger\right)_{\mu\mu} \right|^2  = s^{4}_{23}|\mathcal{T}_{\tau\tau}|^{2} + c^{4}_{23}|\mathcal{T}_{\mu\mu}|^{2}+2c^2_{23}s^2_{23}\Re\{\mathcal{T}_{\mu \mu}^\ast \mathcal{T}_{\tau \tau}\} &  \nonumber \\
 &=s^{4}_{23} \left[1-\sin^22\widetilde{\theta}_{13}\sin^2\left(\dfrac{\Delta\widetilde{m}^2_{31}}{4E_{\nu}}L\right)\right]
+c^{4}_{23}+& \nonumber \\
 & 2c^2_{23}s^2_{23}\left[\cos\left(\dfrac{\Delta\widetilde{m}^2_{31}}{4E_\nu}L\right)\cos\left(\dfrac{\Delta m_{31}^{2,{\rm eff}}+2E_\nu V}{4E_\nu}L\right)+\cos2\widetilde{\theta}_{13}\sin\left(\dfrac{\Delta\widetilde{m}^2_{31}}{4E_\nu}L\right)\sin\left(\dfrac{\Delta m_{31}^{2,{\rm eff}}+2E_\nu V}{4E_\nu}L\right)\right]~, &
\label{full-Pvep}
\end{align}
where $\mathcal{T}$ is the matrix in Eq.~(\ref{smatrix-casei}). Below the resonance energy, $E_\nu^{{\rm res},31}$, the $\sin2\widetilde{\theta}_{13}$ is suppressed and the following simple relation recovers
\begin{equation}
P(\nu_{\mu}\to \nu_{\mu})=1-\sin^{2}2\theta_{23}\sin^{2}\left(\dfrac{\Delta m^{2,{\rm eff}}_{31}}{4E_{\nu}}L    \right)~.
\label{Pvep}
\end{equation}
At the resonance $\widetilde{\theta}_{13}$ is maximal and the first term in Eq.~(\ref{full-Pvep}) leads to $\nu_\mu\to\nu_e$ conversion in the resonance region. Above the resonance region $\widetilde{\theta}_{13}\to\theta_{13}$. However, since $\theta_{13}$ is small ($s_{13}^2\simeq0.02$) the $\nu_\mu\to\nu_e$ oscillation in the high energy range is quite small. Thus, in summary, the oscillation probability in Eq.~(\ref{Pvep}) is a good approximation of $\nu_\mu$ survival probability for $\phi\Delta\gamma_{31}\neq0$, except for the very narrow resonance region where $\nu_\mu\to\nu_e$ oscillation exists.

The same condition applied in Eqs.~(\ref{emin}) and (\ref{emax}) to find the minima and maxima of $\nu_\mu$ survival probability can be applied to Eq.~(\ref{Pvep}), which leads to a quadratic equation for $E_\nu$. The minima of $\nu_\mu$ survival probability in Eq.~(\ref{Pvep}) are at $E_{\nu,{\rm VEP}}^{{\rm min},n}$ which stems from the condition
\begin{equation}
\frac{\Delta m^{2,{\rm eff}}_{31}L}{4E_{\nu,{\rm VEP}}^{{\rm min},n}}=\left(n +\frac{1}{2}\right) \pi \quad\longrightarrow\quad 
a(E_{\nu,{\rm VEP}}^{{\rm min},n})^{2}+bE_{\nu,{\rm VEP}}^{{\rm min},n} + c = 0~,
\label{eq:bask01}
\end{equation}
where, the coefficients of quadratic equation are
\begin{align}
a= \frac{L \phi  \Delta\gamma_{31}}{(n+1/2)\pi}\quad ,  \quad b= -1\quad , \quad 
c= \frac{\Delta m^{2}_{31}L}{4(n+1/2)\pi}~.
\label{eq:abc}
\end{align}
Solutions of the quadratic equation in Eq.~(\ref{eq:bask01}) gives the minima of the $\nu_\mu$ survival probability $E_{\nu,{\rm VEP}}^{{\rm min},n}$ in the presence VEP as ($n=0,1,\ldots$)
\begin{equation}\label{eq:vepmin}
E_{\nu,{\rm VEP}}^{{\rm min},n} = \frac{\left(n+1/2 \right)\pi}{2L\phi\Delta\gamma_{31}} \left[ 1\pm \sqrt{1- \frac{\Delta m_{31}^2L^2\phi\Delta\gamma_{31}}{\left[ \left( n+1/2\right)\pi\right]^2}}~ \right]~.
\end{equation}
For $\phi\Delta\gamma_{31}\lesssim10^{-25}$ the second term inside the square root in Eq.~(\ref{eq:vepmin}) is small even for the largest propagation length ($L=2R_\oplus$) and $n=0$. Thus, using the approximation $\sqrt{1-x}=1-x/2$, two sets of solution for $E_{\nu,{\rm VEP}}^{{\rm min},n}$ can be obtained: the first set is equal to minima in standard oscillation, $E_{\nu,{\rm std}}^{{\rm min},n}$ in Eq.~(\ref{emin}), and the second set is  
\begin{equation}\label{eq:vepeminexact}
E_{\nu,{\rm VEP}}^{{\rm min},n} = \frac{\left( n+ 1/2\right)\pi}{L\phi\Delta\gamma_{31}} - E_{\nu,{\rm std}}^{{\rm min},n}~.
\end{equation}
The first set of solutions gives the conventional minima in low energy range ($\lesssim 20$~GeV) while the second set of solutions introduce new minima in the high energy range. Since the maximum value of the second term in Eq.~(\ref{eq:vepeminexact}) is $\sim25$~GeV, the minima in the high energy range are ($n=0,1,\ldots$)
\begin{equation}\label{eq:vepemin}
E_{\nu,{\rm VEP}}^{{\rm min},n} \simeq \frac{\left( n+ 1/2\right)\pi}{L\phi\Delta\gamma_{31}} = 2.43~ {\rm TeV}\, \left( \frac{2n+1}{1}\right) \left( \frac{-1}{\cos\theta_z}\right) \left( \frac{10^{-26}}{\phi\Delta\gamma_{31}}\right)~.
\end{equation}    
In the same way, the maxima of $\nu_\mu$ survival probability can be obtained by the condition $\Delta m^{2,{\rm eff}}_{31}L/(4E_{\nu,{\rm VEP}}^{{\rm max},k})=k\pi$; which again leads to two sets of solutions: one set equal to $E_{\nu,{\rm std}}^{{\rm max},k}$ in Eq.~(\ref{emax}) and the second set given by ($k=1,2,\ldots$)
\begin{equation}\label{eq:vepemax}
E_{\nu,{\rm VEP}}^{{\rm max},k} \simeq \frac{k\pi}{L\phi\Delta\gamma_{31}} = 4.86~ {\rm TeV}\, \left( \frac{k}{1} \right) \left( \frac{-1}{\cos\theta_z}\right) \left( \frac{10^{-26}}{\phi\Delta\gamma_{31}}\right)~.
\end{equation}

\begin{figure}[t!]
\includegraphics[scale=1]{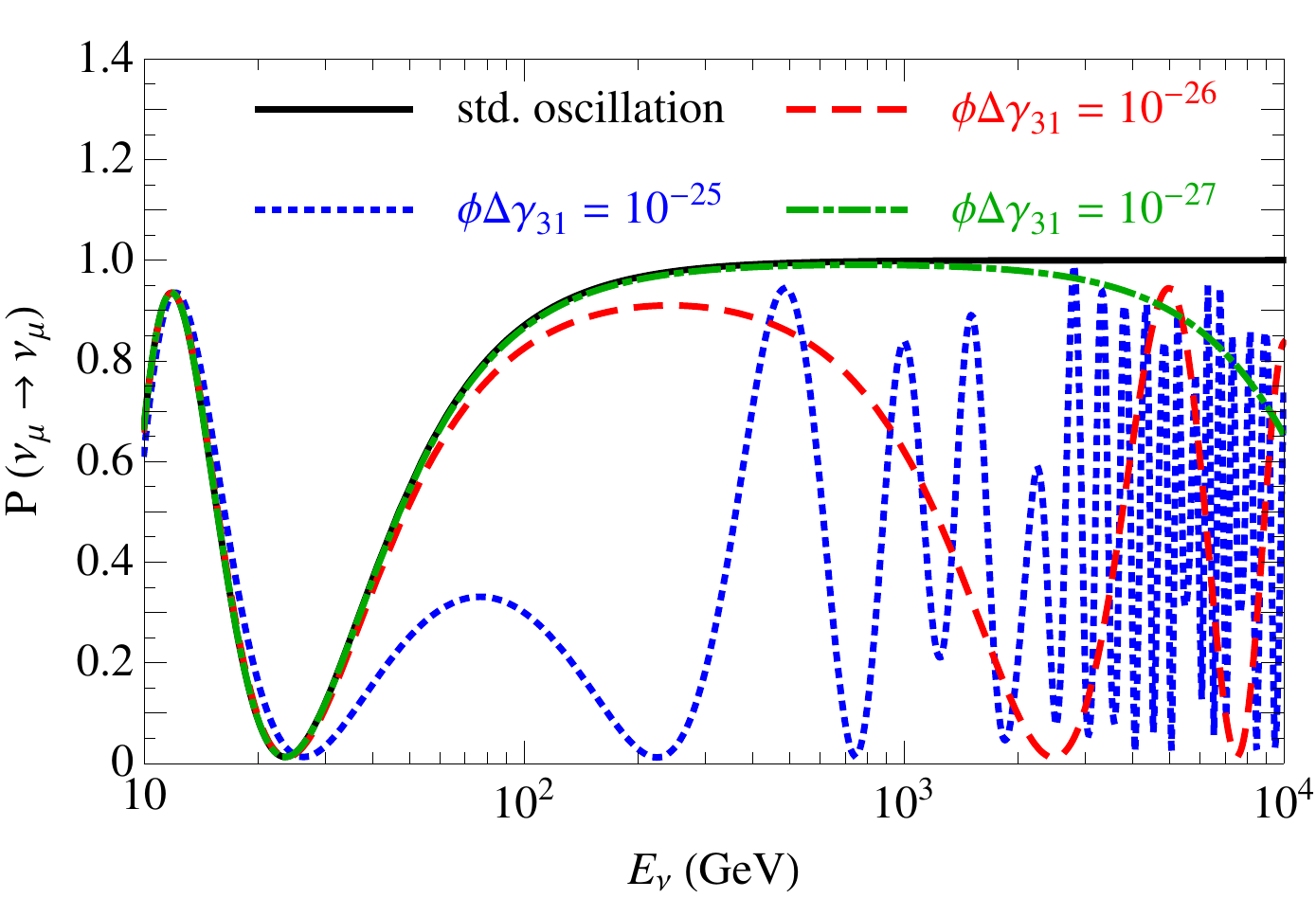}
\caption{\label{fig:pmumu_vep+_cz1}The $\nu_\mu$ survival probability as function of neutrino energy for the Standard Oscillations (solid line) and for VEP scenario (\textbf{Case \textit{i}}) with different values of $\phi\Delta\gamma_{31}$. All the curves are for $\cos\theta_z=-1$. The mixing parameters are fixed at best-fit values from~\cite{GonzalezGarcia:2012sz}.}
\end{figure}

Thus, in summary, in the presence of VEP with $\phi\Delta\gamma_{31}\neq0$ in addition to the conventional minima and maxima in $\nu_\mu$ survival probability in the low energy range ($\lesssim30$~GeV), a new set of maxima and minima exists which, for $\phi\Delta \gamma_{31}\lesssim10^{-25}$, appear in the high energy range ($\gtrsim100$~GeV). So, in the presence of VEP with $\phi\Delta \gamma_{31}\lesssim10^{-25}$, although the phenomenology of low energy atmospheric neutrinos do not change, the high energy range drastically modifies by the new minima and maxima. A feature of minima and maxima energies, in Eqs.~(\ref{eq:vepemin}) and (\ref{eq:vepemax}) respectively, worths to mention: although for the standard oscillation the energies of minima and maxima decrease with the increase of $\cos\theta_z$ (see Eqs.~(\ref{emin}) and (\ref{emax})), in the presence of VEP the minima and maxima energies in Eqs.~(\ref{eq:vepemin}) and (\ref{eq:vepemax}) increase with the increase of $\cos\theta_z$.

In Fig.~\ref{fig:pmumu_vep+_cz1} we show numerical calculation of the $\nu_\mu$ survival probability in the presence of VEP with $\phi\Delta\gamma_{31}=10^{-25}$, $10^{-26}$ and $10^{-27}$, by blue dotted, red dashed and green dot-dashed curves, respectively. Also, the black solid line show the standard $\nu_\mu$ survival probability. All the curves in Fig.~\ref{fig:pmumu_vep+_cz1} are for $\cos\theta_z=-1$. For the blue dotted curve, where $\phi\Delta\gamma_{31}=10^{-25}$, the resonance at $\sim1.8$~TeV can be identified. For smaller values of $\phi\Delta\gamma_{31}$ resonance is out of the range of Fig.~\ref{fig:pmumu_vep+_cz1}.  As it can be seen, the pattern of minima of maxima are in agreement with Eqs.~(\ref{eq:vepemin}) and (\ref{eq:vepemax}). For example, for $\phi\Delta\gamma_{31}=10^{-26}$, in addition to conventional minima and maxima in $E_\nu\lesssim30$~GeV, we expect the first minimum and maximum at $E_{\nu,{\rm VEP}}^{{\rm min},0}=2.43$~TeV and $E_{\nu,{\rm VEP}}^{{\rm max},1}=4.86$~TeV respectively; which are clearly visible in red dashed curve of Fig.~\ref{fig:pmumu_vep+_cz1}. For larger values of $\phi\Delta\gamma_{31}$ the minima/maxima push to lower energies. For $\phi\Delta\gamma_{31}=10^{-25}$ the first minimum and maximum would be at $E_{\nu,{\rm VEP}}^{{\rm min},0}\simeq220$~GeV (from Eq.~(\ref{eq:vepeminexact})) and $E_{\nu,{\rm VEP}}^{{\rm max},1}\simeq486$~GeV respectively, which are in agreement with the blue dotted curve. On the other hand, for smaller values of $\phi\Delta\gamma_{31}$ the substantial deviation of $\nu_\mu$ survival probability from standard pattern occur at higher energies. For $\phi\Delta\gamma_{31}=10^{-27}$ the first minimum is at $E_{\nu,{\rm VEP}}^{{\rm min},0}\simeq24$~TeV (which is out of the plotted range in Fig.~\ref{fig:pmumu_vep+_cz1}, see the green dot-dashed curve). Although this deviation is in the energy range of IC-40 data set, due to small statistics at thigh energy, sensitivity to these small values of $\phi\Delta\gamma_{31}$ will be quite challenging.

\subsection{Case \textbf{\textit{ii}} : $\phi\Delta\gamma_{21}\neq0$ and $\phi\Delta\gamma_{31}=0$}
\label{sub:caseii}

When $\phi\Delta\gamma_{31}=0$, the VEP do not change 13-mass squared difference while 12-mass squared differences will be modified. However, as we pointed already, in the energy range we are considering the contribution from $\Delta m^{2}_{21}$ can be neglected and so Eq.~(\ref{eq:Dm2eff}) contains only the contribution from VEP. Thus, the effective mass squared differences are:
\begin{equation}
\Delta m^{2,{\rm eff}}_{21}=4E^{2}_{\nu}\phi\Delta\gamma_{21}~~~~,~~~\Delta m^{2,{\rm eff}}_{31}=\Delta m^{2}_{31}~.
\label{Dm2effi}
\end{equation}
In the analytical discussions of this section we assume $\Delta\gamma_{21}>0$, unless mentioned otherwise. Generalization to $\Delta\gamma_{21}<0$ is straightforward. Since $\Delta m^{2,{\rm eff}}_{21}$ increases with energy, there is no decoupling of $\nu_{e}$ from $\nu_{\mu/\tau}$ anymore and the full $3\nu$ system would be considered. In the basis $| \nu^{\prime\prime}\rangle= U^{\dagger}_{13}U^{\dagger}_{23}| \nu \rangle$, the evolution equation, Eq.~(\ref{Sch}), can be written as
\begin{eqnarray}
i\dfrac{{\rm d}}{{\rm d}t}
\left(
\begin{array}{cc}
\nu_{e}^{\prime\prime} \\
\nu_{\mu}^{\prime\prime}\\
\nu_{\tau}^{\prime\prime}\\ 
\end{array}
\right)
=\left(
\begin{array}{ccc}
2s^{2}_{12}E_{\nu}\phi\Delta\gamma_{21} + c^{2}_{13}V & 2s_{12}c_{12}E_{\nu}\phi\Delta\gamma_{21} & c_{13}s_{13}V \\
2s_{12}c_{12}E_{\nu}\phi\Delta\gamma_{21} & 2c^{2}_{12}E_{\nu}\phi\Delta\gamma_{21} & 0 \\
c_{13}s_{13}V & 0 & \dfrac{\Delta m_{31}^2}{2E_\nu}+s^{2}_{13}V \\ 
\end{array}
\right)
\left(
\begin{array}{cc}
\nu_{e}^{\prime\prime} \\
\nu_{\mu}^{\prime\prime}\\
\nu_{\tau}^{\prime\prime}\\ 
\end{array}
\right)~.
\label{rotate}
\end{eqnarray}
Notice that by neglecting the terms proportional to $s_{13}$ the matrix in Eq.~(\ref{rotate}) is block-diagonal and so $|\nu^{\prime\prime}_{\tau} \rangle$ decouple from the rest of states. In this case, the evolution matrix for constant density is
\begin{align}
\left(
\begin{array}{cc}
\nu_{e}^{\prime\prime} \\
\nu_{\mu}^{\prime\prime} \\
\nu_{\tau}^{\prime\prime} \\ 
\end{array}
\right)_{t=L}
=\left(
\begin{array}{ccc}
\mathcal{S}_{ee}       & \mathcal{S}_{e\mu}     & 0 \\
\mathcal{S}_{\mu e} & \mathcal{S}_{\mu\mu} & 0 \\ 
0             & 0                   & \mathcal{S}_{\tau\tau}
\end{array}
\right)
\left(
\begin{array}{cc}
\nu_{e}^{\prime\prime} \\
\nu_{\mu}^{\prime\prime} \\
\nu_{\tau}^{\prime\prime} \\ 
\end{array}
\right)_{t=0}~,
\label{smatrix}
\end{align}
where $\mathcal{S}_{\alpha\beta}$ are
\begin{equation}
\mathcal{S}_{ee,\mu\mu} =\cos\left( \dfrac{\Delta\widetilde{m}^2_{21}}{4E_\nu}L\right)\mp i \cos 2\widetilde{\theta}_{12}\sin\left(\dfrac{\Delta\widetilde{m}^2_{21}}{4E_\nu}L\right) \quad ; \quad \mathcal{S}_{e\mu}=\mathcal{S}_{\mu e}=-i \sin 2\widetilde{\theta}_{12}\sin\left(\dfrac{\Delta\widetilde{m}^2_{21}}{4E_\nu}L\right)~,
\label{Dmeff21}
\end{equation}
and
\begin{equation}
\Delta \widetilde{m}^{2}_{21}=4E_\nu\sqrt{(\cos2\theta_{12}E_\nu\phi \Delta\gamma_{21}-V/2)^2+(\sin2\theta_{12}E_\nu\phi\Delta \gamma_{21})^2} \quad , \quad  
\sin 2\widetilde{\theta}_{12}=\sin 2 \theta_{12}\dfrac{4E_{\nu}^2\phi\Delta \gamma_{21} }{\Delta\widetilde{m}^2_{21}}~,
\label{c2tms2tm}
\end{equation}
are the effective mass-squared difference and 12-mixing angle in matter in $|\nu^{\prime\prime}_\alpha\rangle$ basis.
Also,
\begin{equation}
\mathcal{S}_{\tau\tau}= \exp{\left[-i\dfrac{\Delta \eta}{2E_\nu}L\right]}~,
\label{Dmef31}
\end{equation}
where $\Delta\eta\equiv\Delta m_{31}^2 -E_\nu V - 2E_\nu^2\phi\Delta\gamma_{21}$, is the vacuum amplitude for the decoupled $|\nu^{\prime\prime}_{\tau} \rangle$ state. The $\nu_\mu$ survival probability is given by (neglecting terms proportional to $s_{13}$)
\begin{align}
 P (\nu_{\mu}\to\nu_{\mu}) &= \left| \left(U_{23} \mathcal{S} U_{23}^{\dagger}\right)_{\mu\mu} \right|^2  = c^{4}_{23}|\mathcal{S}_{\mu\mu}|^{2}+s^{4}_{23}|\mathcal{S}_{\tau\tau}|^{2}+2c^2_{23}s^2_{23}\Re\{\mathcal{S}_{\mu \mu}^\ast\mathcal{S}_{\tau \tau}\}+{\cal{O}}(s_{13}) &  \nonumber \\
 &=c^{4}_{23} \left[1-\sin^22\widetilde{\theta}_{12}\sin^2\left(\dfrac{\Delta\widetilde{m}^2_{21}}{4E_{\nu}}L\right)\right]
+s^{4}_{23}+& \nonumber \\
 & 2c^2_{23}s^2_{23}\left[\cos\left(\dfrac{\Delta\widetilde{m}^2_{21}}{4E_\nu}L\right)\cos\left(\dfrac{\Delta\eta}{2E_\nu}L\right)-\cos2\widetilde{\theta}_{12}\sin\left(\dfrac{\Delta\widetilde{m}^2_{21}}{4E_\nu}L\right)\sin\left(\dfrac{\Delta\eta}{2E_\nu}L\right)\right]~, &
\label{pmurotate}
\end{align}
where $\mathcal{S}$ is the matrix in Eq.~(\ref{smatrix}). For $\phi\Delta\gamma_{21}\to 0$ the oscillation probability in the standard scenario will be recovered.

\begin{figure}[t!]
\includegraphics[scale=0.34,angle=0]{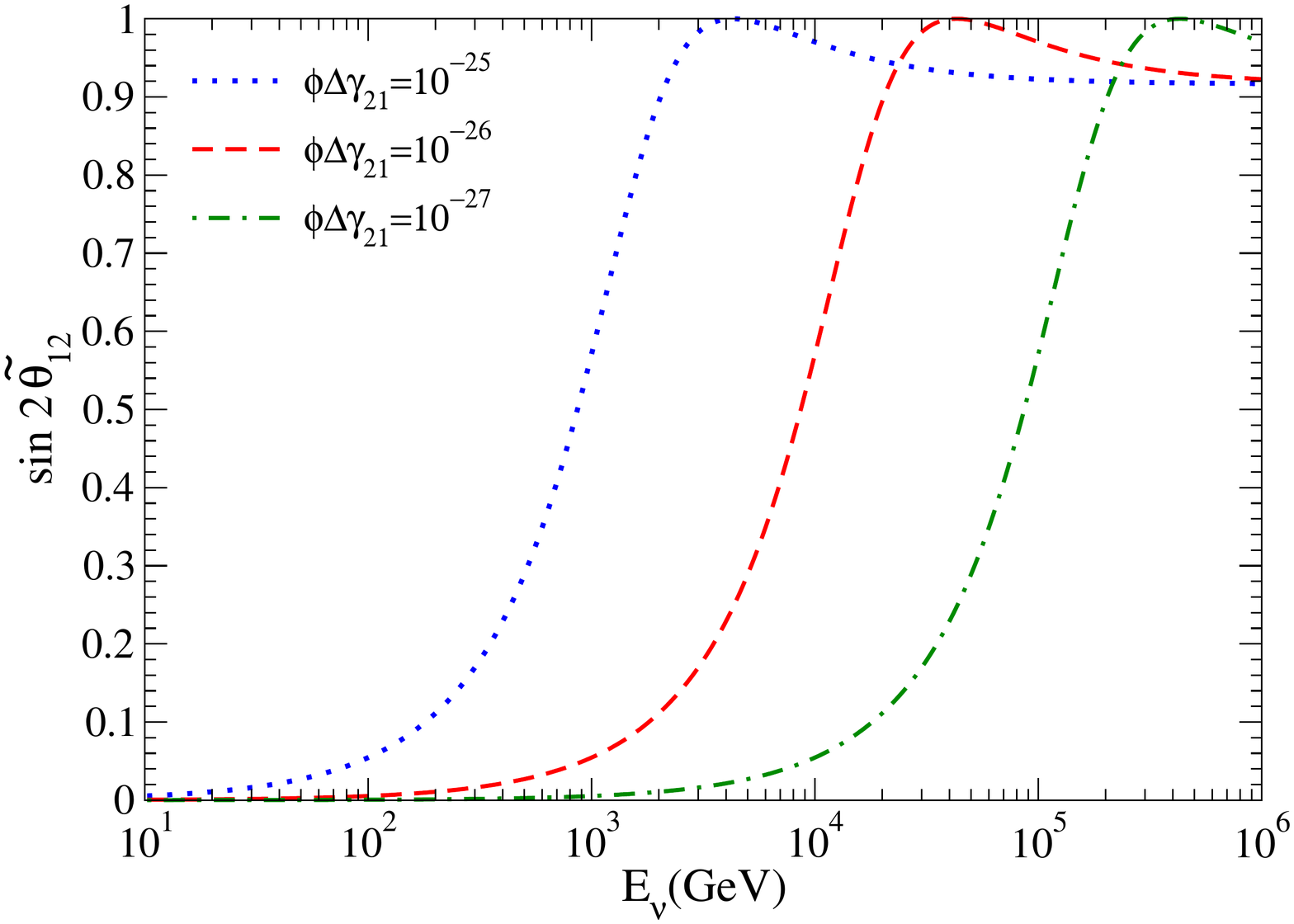}\includegraphics[scale=0.34,angle=0]{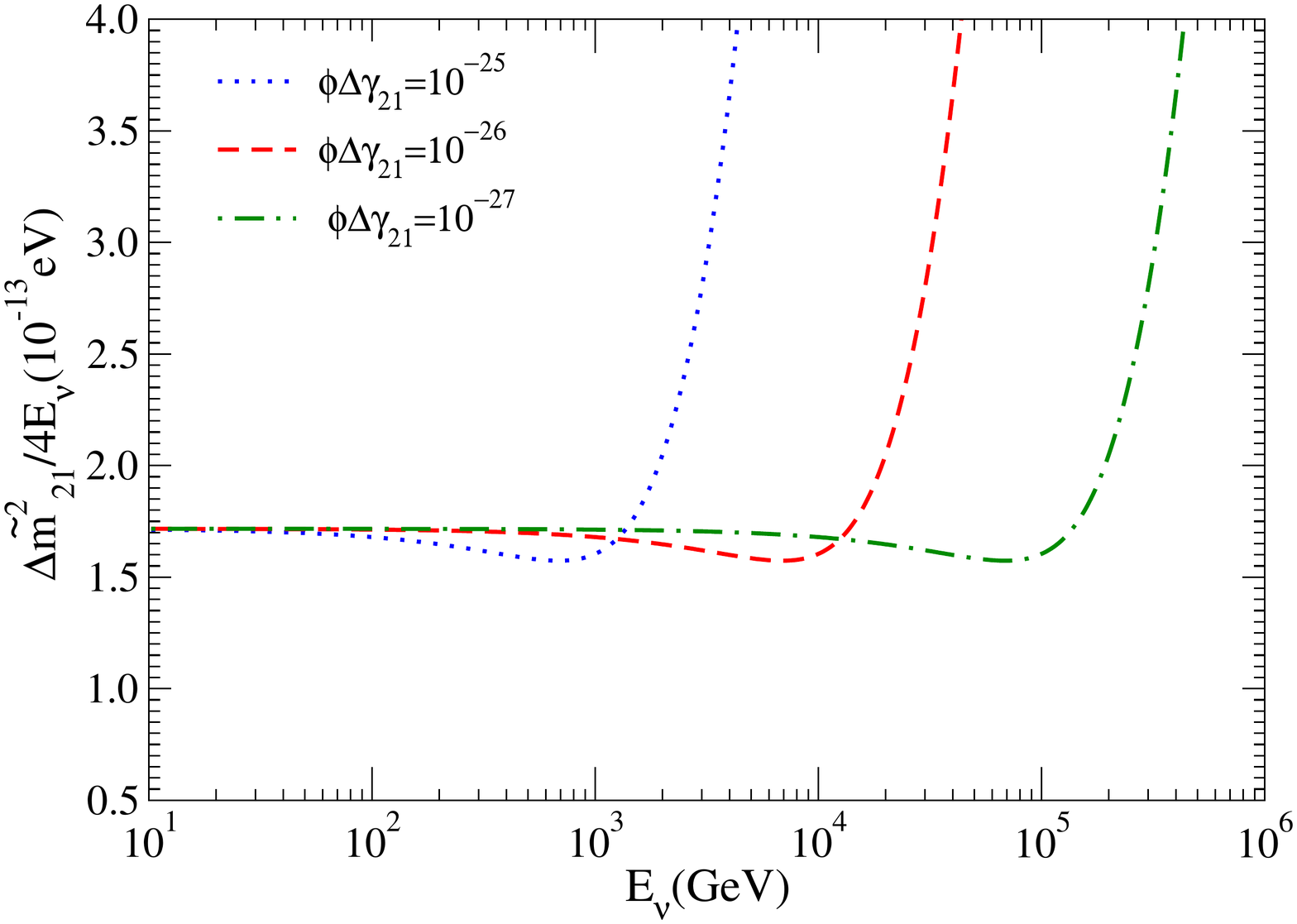}
\caption{\label{fig:q12matter}The dependence of $\sin2\widetilde{\theta}_{12}$ and $\Delta \widetilde{m}^{2}_{21}$ on energy for different values of VEP parameter $\phi\Delta\gamma_{21}$, in the left and right panel respectively. The mixing parameters are set to the best-fit values in~\cite{GonzalezGarcia:2012sz}.}
\end{figure}

The following comments about the oscillation probability in Eq.~(\ref{pmurotate}) are in order. The first term in Eq.~(\ref{pmurotate}), inside the bracket proportional to $c_{23}^4$, is the contribution of $\nu_\mu\to\nu_e$ to $\nu_\mu$ survival probability and in the limit $\phi\Delta\gamma_{21}\to0$ it goes to one; {\it i.e.}, $P(\nu_\mu\to\nu_e)=0$. But, for $\phi\Delta\gamma_{21}\neq0$ this term leads to $\nu_\mu\leftrightarrow\nu_e$ oscillation in the high energy range. This term is regulated by the $\nu_\mu-\nu_\tau$ mixing. The last term in Eq.~(\ref{pmurotate}) is the interference of VEP and $\Delta m_{31}^2$-induced oscillations. From Eq.~(\ref{c2tms2tm}), for $\phi\Delta\gamma_{21}>0$ obviously there is a resonance in neutrino channel at the energy
 \begin{equation}
E_\nu^{{\rm res},21}=\dfrac{V}{2\phi\Delta\gamma_{21}\cos 2\theta_{12}}\simeq 42~{\rm TeV} \left(\dfrac{10^{-26}}{\phi\Delta\gamma_{21}}\right) \left(\dfrac{0.4}{\cos2\theta_{12}}\right) \left(\dfrac{\langle \rho Y_e \rangle}{4.5\,{\rm gcm}^{-3}}\right)~.
\label{echar}
\end{equation}
The resonance is in the antineutrino channel for $\phi\Delta\gamma_{21}<0$. In the left and right plot of Fig.~\ref{fig:q12matter} we show $\sin2\widetilde{\theta}_{12}$ and $\Delta\widetilde{m}_{21}^2$, respectively, as a function of energy for various values of $\phi\Delta\gamma_{21}$. The resonance at $E_\nu^{{\rm res},21}$ can be clearly identified as maximum in $\sin2\widetilde{\theta}_{12}$ and minimum in $\Delta\widetilde{m}_{21}^2$ values. At the resonance, $E_\nu\sim E_\nu^{{\rm res},21}$, the effective angle $\widetilde{\theta}_{12}$ is maximal and a complete conversion of $\nu_\mu\to\nu_{e/\tau}$ occurs. At energies higher than resonance energy, $E_\nu\gtrsim E_{\nu}^{{\rm res},21}$, vacuum oscillation recovers: the effective mixing angle approaches the vacuum value, $\widetilde{\theta}_{12} \to \theta_{12}$, and oscillation is governed by the effective 21-mass squared difference $\Delta \widetilde{m}^{2}_{21}=4E_\nu^2\phi\Delta\gamma_{21}$ which induce $\nu_\mu\to\nu_{e/\tau}$ oscillation. However, since $\Delta \widetilde{m}^{2}_{21} \propto E_\nu^2$, the mass-squared difference is large which leads to fast oscillatory behavior at $E_\nu\gtrsim E_{\nu}^{\rm res}$. Below the resonance energy, $E_\nu\lesssim E_{\nu}^{{\rm res},21}$, the effective mixing $\widetilde{\theta}_{12}$ is suppressed and so $P(\nu_\mu\to\nu_e)\simeq 0$. So, for $E_\nu\lesssim E_\nu^{{\rm res},21}$ where $\sin 2\widetilde{\theta}_{12}\simeq0$, the $\nu_\mu$ survival probability in Eq.~(\ref{pmurotate}) reduces to
\begin{equation}
P(\nu_{\mu}\to\nu_{\mu})\simeq 1-\sin^22\theta_{23}\sin^2\left[\dfrac{\left(\Delta m^2_{31} - 2E_\nu^2\phi\Delta\gamma_{21}\right)}{4E_\nu}L\right]~.
\label{approx-caseii}
\end{equation}
The minima and maxima of Eq.~(\ref{approx-caseii}) can be obtained in the same way as \textbf{Case \textit{i}} discussed in section~\ref{sub:casei}~. With a straightforward calculation it can be shown that, neglecting the contribution of $\Delta m_{31}^2$, the minima and maxima of Eq.~(\ref{approx-caseii}) are at $2E_{\nu,{\rm VEP}}^{{\rm min},n}$ and $2E_{\nu,{\rm VEP}}^{{\rm max},k}$ respectively (see Eqs.~(\ref{eq:vepemin}) and (\ref{eq:vepemax})). However, these minima (maxima) with depth (height) controlled by $\sin^22\theta_{23}$ exist when $2E_{\nu,{\rm VEP}}^{{\rm min},n}\lesssim E_\nu^{{\rm res},21}$ ($2E_{\nu,{\rm VEP}}^{{\rm max},k}\lesssim E_\nu^{{\rm res},21}$).  At higher energies the oscillation is induced by $\Delta\widetilde{m}_{21}^2$ and $\sin^22\theta_{12}$.

\begin{figure}[t!]
\includegraphics[angle=0,scale=1]{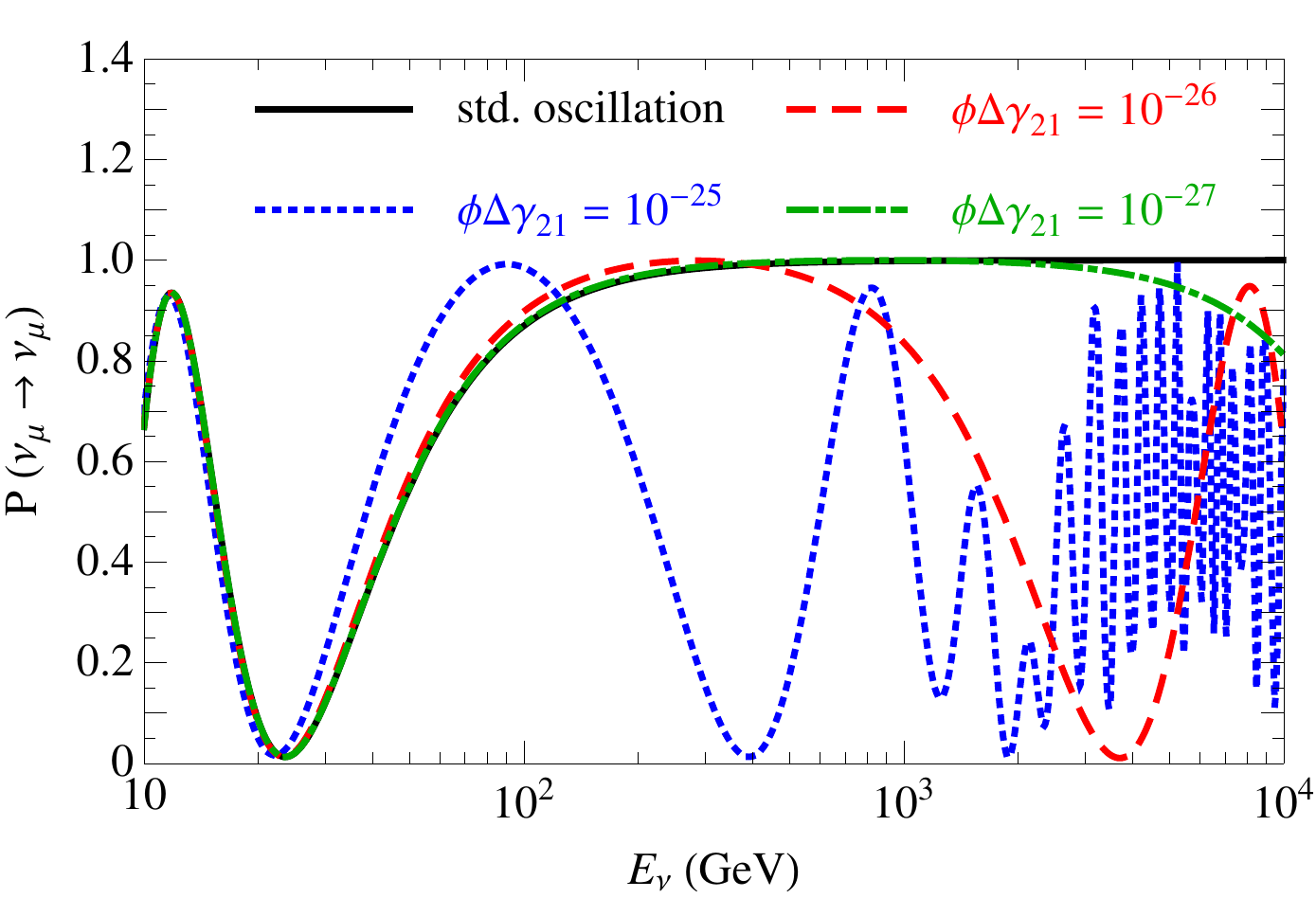}
\caption{\label{fig:pmumucasei}The $\nu_\mu$ survival probability as function of neutrino energy for the Standard Oscillations (solid line) and for VEP scenario (\textbf{Case \textit{ii}}) with different values of $\phi\Delta\gamma_{21}$. All the curves are for $\cos\theta_z=-1$. The mixing parameters are fixed at best-fit values from~\cite{GonzalezGarcia:2012sz}.}
\end{figure}

In Fig.~\ref{fig:pmumucasei} we show numerical calculation of the $\nu_\mu$ survival probability for different values $\phi\Delta\gamma_{21}$, for $\cos\theta_z=-1$. The features discussed above are manifest: for example, taking $\phi\Delta\gamma_{21}=10^{-25}$ which is shown by the blue dotted curve in Fig.~\ref{fig:pmumucasei}, the resonance at $\sim4$~TeV can be seen. Above the resonance oscillation is induced by $\Delta \widetilde{m}^{2}_{21}$ which lead to the fast oscillatory behavior. In lower energies, oscillation is governed by $\Delta \widetilde{m}^{2}_{21}$ and $\sin^22\theta_{23}$, which leads to minima and maxima with the double energies with respect to the \textbf{Case \textit{i}} in section~\ref{sub:casei} (compare with the position of minima and maxima in Fig.~\ref{fig:pmumu_vep+_cz1}). The oscillation probability derived in Eq.~(\ref{approx-caseii}) is in good agreement with the numerical result shown in Fig.~\ref{fig:pmumucasei}. 

Comparing the curves in Fig.~\ref{fig:pmumucasei} with standard oscillation (black solid curve) shows that for $\phi\Delta\gamma_{21}\lesssim10^{-27}$ the effect of VEP shifts to energies higher than $\sim10$~TeV, where the flux of atmospheric neutrinos is very small. Thus, neutrino telescopes are sensitive to $\phi\Delta\gamma_{21}\gtrsim10^{-27}$.

\subsection{Case \textbf{\textit{iii}} : $\phi\Delta\gamma_{21}=\phi\Delta\gamma_{31}\neq0$} 
\label{sub:caseiii}

When $\phi \Delta\gamma_{21} =\phi \Delta\gamma_{31}\equiv\phi\Delta \gamma$, after subtracting $\phi\Delta\gamma\mathbf{I}$ (where $\mathbf{I}$ is the unit matrix), the coupling matrix of neutrinos to gravitational field is $\Delta G={\rm diag}(-4E_\nu^4\phi\Delta\gamma,0,0)$. The oscillation probabilities can be calculated in a similar way as in section~\ref{sub:caseii}: by changing the basis to $|\nu_\alpha^{\prime\prime}\rangle$, the Hamiltonian takes a block-diagonal form. By straightforward calculation, it can be shown that the $\nu_\mu$ survival probability is similar to Eq.~(\ref{pmurotate}) with the replacement $\Delta\eta\to\Delta\eta^\prime = \Delta m_{31}^2 - E_\nu V+2E_\nu^2\phi\Delta\gamma$. Thus, all the discussions of section~\ref{sub:caseii} applies here, including the resonance and $\nu_\mu\to\nu_e$ conversion at high energies, with the exception that the set of minima and maxima below the resonance is absent here, mainly since the minima and maxima of \textbf{Case \textit{i}} and \textbf{Case \textit{ii}} interfere and cancel each other. This absence of oscillatory behavior below the resonance energy, which means less deviation from standard oscillation, leads to weaker limit on VEP parameters when $\phi \Delta\gamma_{21} =\phi \Delta\gamma_{31}$, as we show in section~\ref{analysis}.

\subsection{Oscillograms}
\label{Oscillograms}

In the previous subsections we discussed analytically the main features induced by VEP on atmospheric neutrino oscillation in the high energy range. Also we showed the numerical calculation of $\nu_\mu$ survival probability for up-going neutrinos at IceCube; {\it i.e.}, neutrinos which pass the diameter of Earth and so their incoming direction have $\cos\theta_z=-1$. In this section we present the oscillograms of $\nu_\mu$ survival probability which illustrates, among the others, also the zenith dependence of probability.

\begin{figure}[!ht]
\centering
\subfloat[standard oscillation]{
 \includegraphics[trim= 0mm 0mm 0mm 
95mm,clip,width=0.45\textwidth]{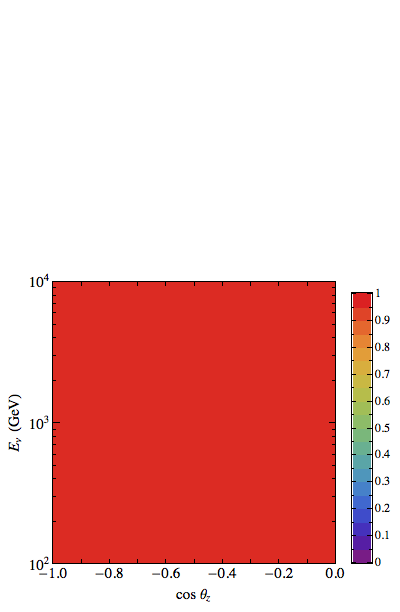}
  \label{fig:so}
}
\subfloat[$\phi\Delta \gamma_{21}=0$ and $\phi\Delta \gamma_{31}=10^{-26}$]{
 \includegraphics[trim= 0mm 0mm 0mm 
95mm,clip,width=0.45\textwidth]{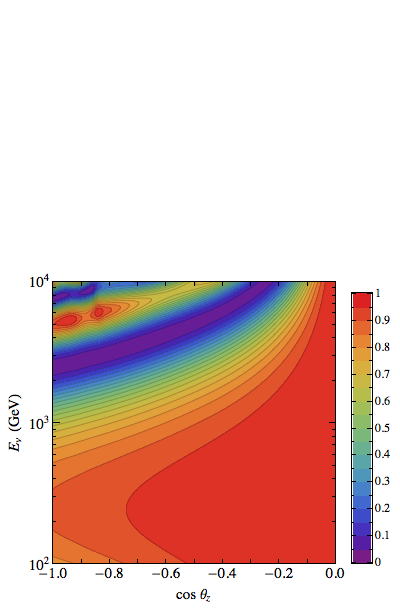}
  \label{fig:casei}
}
\quad
\subfloat[$\phi\Delta \gamma_{21}=10^{-26}$ and $\phi\Delta \gamma_{31}=0$]{
 \includegraphics[trim= 0mm 0mm 0mm 
95mm,clip,width=0.45\textwidth]{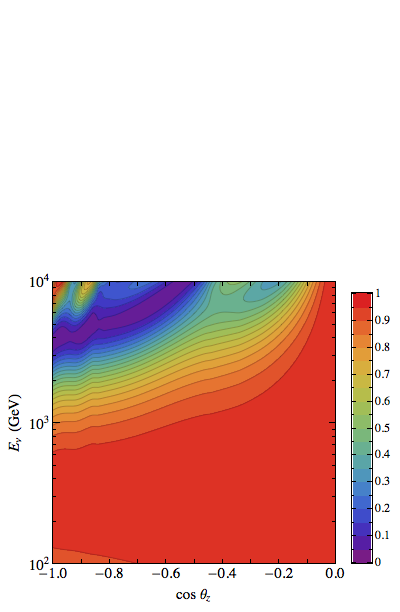}
  \label{fig:caseii}
}
\subfloat[$\phi\Delta \gamma_{21}=\phi\Delta \gamma_{31}=10^{-26}$]{
 \includegraphics[trim= 0mm 0mm 0mm 
95mm,clip,width=0.45\textwidth]{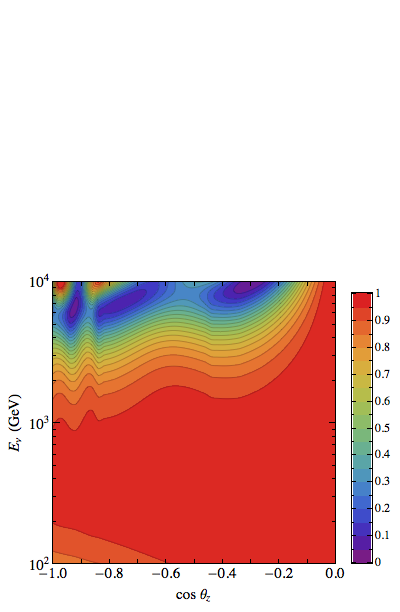}
  \label{fig:caseiii}
}
\caption{\label{fig:pmumu_vep+} The oscillograms for survival probability $P(\nu_\mu\to\nu_\mu)$. The top-left, top-right, bottom-left and bottom-right panels are respectively for standard oscillation, \textbf{Case \textit{i}}, \textbf{Case \textit{ii}} and \textbf{Case \textit{iii}}. The values of $\phi\Delta\gamma_{ij}$ are indicated in the captions. The mixing parameters are fixed to their best-fit values from~\cite{GonzalezGarcia:2012sz}.}
\end{figure}

Fig.~\ref{fig:pmumu_vep+} shows the oscillograms of the $\nu_\mu$ survival probability. The panel~\ref{fig:so} is for the standard oscillation; {\it i.e.}, $\phi\Delta\gamma_{21}=\phi\Delta\gamma_{31}=0$. As we expect, in the shown energy range $E_\nu>100$~GeV, $P(\nu_\mu\to\nu_\mu)=1$ except for percent-level deviation at $E_\nu\sim100$~GeV and $\cos\theta_z\simeq-1$. In panel~\ref{fig:casei} we show the $\nu_\mu\to\nu_\mu$ oscillation probability for the \textbf{Case \textit{i}} in section~\ref{sub:casei} with $\phi\Delta\gamma_{31}=10^{-26}$. The resonance energy $E_\nu^{{\rm res},31}\simeq18$~TeV is out of the depicted energy range in panel~\ref{fig:casei} and so in all the energy range of this panel $P(\nu_\mu\to\nu_e)\simeq0$. As we discussed in section~\ref{sub:casei}, below the resonance energy, oscillation is almost vacuum oscillation dictated by the VEP effective mass difference, $\Delta m_{31}^{2,{\rm eff}}=\Delta m_{31}^2+2E_{\nu}^2\phi\Delta\gamma_{31}$ and the amplitude $\sin^2 2\theta_{23}$. From Eq.~(\ref{eq:vepemin}) the energy of first VEP-induced minimum in $P(\nu_\mu\to\nu_\mu)$ at $\cos\theta_z=-1$ is $E_{\nu,{\rm VEP}}^{{\rm min},0}\simeq2.4$~TeV, which is in agreement with panel~\ref{fig:casei}. Also, from Eq.~(\ref{eq:vepemin}), the energy of minimum would increase with the increase of $\cos\theta_z$ which is manifest by the violet strip in panel~\ref{fig:casei}.

Panel~\ref{fig:caseii} is for the \textbf{Case \textit{ii}} in section~\ref{sub:caseii} with $\phi\Delta\gamma_{21}=10^{-26}$. In this case also the resonance energy $E_\nu^{{\rm res},21}\simeq42$~TeV is out of the depicted range. From the Eq.~(\ref{approx-caseii}) the first minimum would be at $2E_{\nu,{\rm VEP}}^{{\rm min},0}\simeq4.8$~TeV which is visible in panel~\ref{fig:caseii}. Finally, the panel~\ref{fig:caseiii} is for the \textbf{Case \textit{iii}} with $\phi\Delta\gamma_{21}=\phi\Delta\gamma_{31}=10^{-26}$. As can be seen the pattern of oscillation is similar to the previous two cases, with the exception that in the lower energies the maxima and minima are less profound.

\section{Probing VEP with IceCube Data}
\label{analysis}

In this section we confront the atmospheric neutrino data collected by IceCube with the expectation in the presence of VEP. Generally IceCube can identify two types of events: muon-tracks and cascades. Muon-tracks originate from the charged current interaction of $\nu_\mu$ and $\bar{\nu}_\mu$ which produce respectively $\mu^-$ and $\mu^+$ that their propagation inside the ice emits Cherenkov radiation collectable by photomultipliers implemented in ice\footnote{There is a small contribution to muon-tracks through the charged current interaction of $\nu_\tau$ and $\bar{\nu}_\tau$ and the subsequent leptonic decay of tau particles to muons. This contribution is quite small in the high energy range that we are considering in this paper.}. Cascade events originate from the other interactions including the neutral current interaction of all the neutrino flavors and charged current interaction of $\nu_e$, $\nu_\tau$ and their antineutrinos. Each of these two types of events have advantages and disadvantages: for muon-tracks the benefits are great resolution in reconstruction of the direction of incoming neutrinos and high statistics due to long muon range in ice and rock; while the drawback is the moderate resolution in energy reconstruction. One the other hand, for cascades the energy reconstruction is good while the direction reconstruction of incoming neutrinos is poor. 

In this paper we analyze the atmospheric neutrino data sets IC-40~\cite{Abbasi:2010ie} and IC-79~\cite{Aartsen:2013jza} containing muon-track events collected by respectively 40 and 79 strings out of the final 86 strings of completed IceCube. the energy range of IC-40 ad IC-79 data set are respectively 100~GeV-400~TeV and 100~GeV-10~TeV. To calculate the expected distribution of events in the presence of VEP, we compute numerically the oscillation probabilities by scanning the whole parameter space of $(\phi\Delta\gamma_{21},\phi\Delta\gamma_{31})$. In the numerical computation of probabilities we fix the mixing parameters to their best-fit values from~\cite{GonzalezGarcia:2012sz} and for the density of Earth we use the PREM model~\cite{Dziewonski1981297}. 

Also we calculate the expected sensitivity of IceCube to VEP parameters from cascade events. IceCube already observed atmospheric neutrino induced cascade events with IC-40~\cite{Aartsen:2013vca} and DeepCore~\cite{Aartsen:2012uu} which the latter provided the first measurement of atmospheric $\nu_e$ flux. In section~\ref{cascades} we calculate the the sensitivity of cascade events to VEP parameters assuming the full IceCube detector.

\subsection{Constraints on VEP parameters from IC-40 and IC-79 muon-track data}
\label{Muon-tracks analysis}

The IC-40 and IC-79 data sets are published by IceCube collaboration in 10 bins of $\cos\theta_z$ (from $-1$ to $0$ with bin width $0.1$) and integrated over energy. To analyze these data we calculate the expected number of events in bins of zenith angle and energy, where for the energy bins we takes widths $\Delta \log_{10}(E_\nu/{\rm GeV})=0.3$ and $0.125$, respectively for IC-40 and IC-79 configurations. However, at the end we confront the total number of events in zenith bins (integrated over the neutrino energy) with data. The number of muon-track events in the $i$-th bin of $\cos\theta_z$ and $j$-th bin of $E_\nu$ is given by
\begin{equation}\label{eq:tracks}
N^{\mu}_{i,j}=T\Delta\Omega\left[\sum_{\alpha=e,\mu} \int_{\Delta_i \cos\theta_z} \int_{\Delta_j E_\nu} \Phi_{\nu_{\alpha}}(E_{\nu},\cos \theta_{z}) P(\nu_\alpha\to\nu_\mu)(\{\phi \Delta \gamma_{kl}\})A^{\nu_\mu}_{{\rm eff}}(E_{\nu},\cos \theta_{z})
{\rm d}E_{\nu}{\rm d}\cos\theta_{z} + \nu\to\bar{\nu}\right]~,
\end{equation}
where $T$ is the data-taking time, $\Delta\Omega=2\pi$ is the azimuthal acceptance of IceCube detector, $\Phi_{\nu_\alpha}$ is the atmospheric neutrino flux of $\nu_\alpha$ from~\cite{Honda:2006qj,Athar:2012it}, and $P(\nu_\alpha\to\nu_\mu)$ is the oscillation probability with VEP parameters $\{\phi \Delta \gamma_{kl}\}$. In Eq.~(\ref{eq:tracks}) the $A_{\rm eff}^{\nu_\mu(\bar{\nu}_\mu)}$ is the $\nu_\mu(\bar{\nu}_\mu)$ effective area of IceCube, where for IC-40 is taken from~\cite{Esmaili:2012nz} and for IC-79 have been estimated by rescaling the effective area of IC-40 (the same has been used in~\cite{Esmaili:2013fva,Esmaili:2013vza}).

In the analysis of IC-40 and IC-79 data, we perform a simple $\chi^2$ analysis defined in the following way
\begin{equation}
\chi^{2}( \Delta\gamma_{21}, \Delta\gamma_{31}; \alpha, \beta) = \sum_{i}\left(\dfrac{\left[N^{{\rm data}}_{i}-\alpha(1+\beta(0.5+\cos\theta_{z}))N_{i}^\mu(\phi \Delta \gamma_{21}, \phi  \Delta\gamma_{31})\right]^{2}}{\sigma^{2}_{i,{\rm stat}} + \sigma^{2}_{i,{\rm sys}}} \right)
+ \dfrac{(1-\alpha)^{2}}{\sigma^{2}_{\alpha}}+\dfrac{\beta^{2}}{\sigma^{2}_{\beta}}~,
\end{equation}
where $\sigma_{i,{\rm stat}}=\sqrt{N^{data}_{i}}$ is the statistical error, $\alpha$ and $\beta$ are the parameters that take into account respectively the correlated normalization and zenith dependence uncertainties of the atmospheric neutrino flux with the uncertainties $\sigma_{\alpha}=0.24$ and $\sigma_{\beta}=0.04$~\cite{Honda:2006qj}. The $\sigma_{i,{\rm stat}}=fN_i^\mu$ is the uncorrelated systematic error which for IC-40 and IC-79 we assume $\sim4\%$ and $3\%$ respectively\footnote{The exact values of uncorrelated systematic errors are not reported by IceCube collaboration. We took the mentioned values by requiring statistically meaningful $\chi^2$ values. However, the obtained bounds are quite smooth with respect to changes in the value of $f$.}. The index $i=1,\ldots,10$ runs over zenith bins and $N_i^\mu$ can be obtained from Eq.~(\ref{eq:tracks}) by summing over $j$. After marginalizing with respect to $\alpha$ and $\beta$ pull parameters, upper limit on VEP parameters $(\phi\Delta\gamma_{21},\phi\Delta\gamma_{31})$ can be obtained.

\begin{figure}[!t]
\centering
\includegraphics[trim= 0mm 10mm 0mm 25mm,clip,width=0.8\textwidth,angle=0]{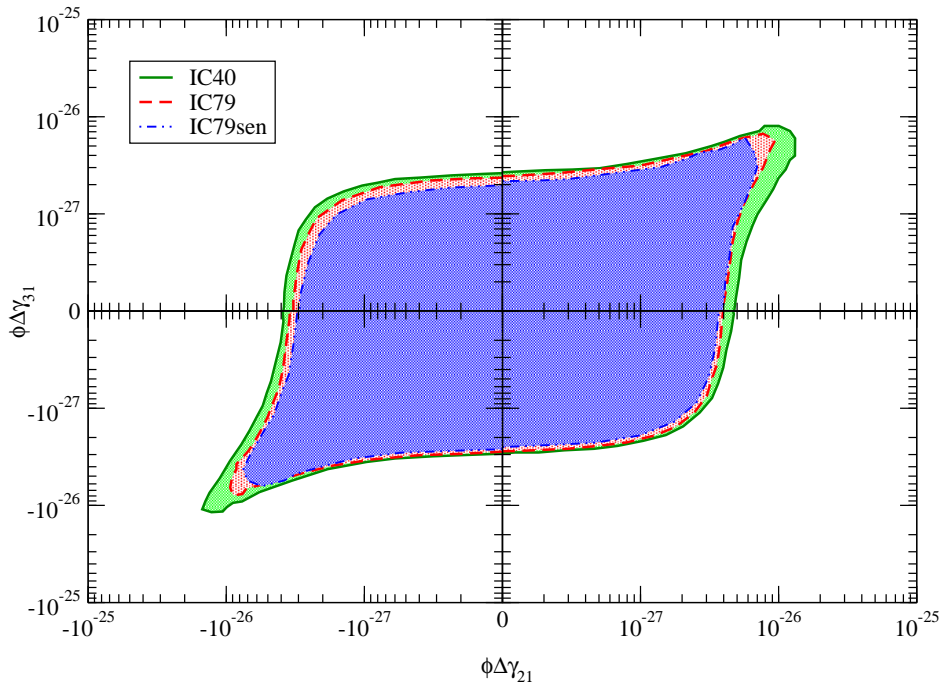}
\caption{\label{fig:limit}The allowed region at 90\% C.L. in the plane $(\phi\Delta\gamma_{21},\phi\Delta\gamma_{31})$. The green solid and red dashed curves are the limits from IC-40 and IC-79 atmospheric muon-track data, respectively. The blue dot-dashed curve shows the sensitivity of IceCube with three times IC-79 data.}
\end{figure}

Fig.~\ref{fig:limit} shows the the obtained limit (at 90\% C.L.) on VEP parameters from analyzing the IC-40 and IC-79 data. In this figure the green solid and red dashed curves are for IC-40 and IC-79 data, respectively. Clearly the weakening of bound at $\phi\Delta\gamma_{21}=\phi\Delta\gamma_{31}$ is visible, which we discussed in section~\ref{sub:caseiii}. The 1-dim limits on VEP parameters are (at 90\% C.L.)
\begin{equation}
-9.2\times10^{-27} < \phi\Delta\gamma_{21} < 9.1\times10^{-27} \quad , \quad  -6.3\times10^{-27} < \phi\Delta\gamma_{31} < 5.6\times10^{-27}~.
\end{equation}
Comparing these limits with the current bounds in Table~\ref{ultima} shows that the limit on $\phi\Delta\gamma_{21}$ is stronger by $\sim4$ orders of magnitude. The limit on $\phi\Delta\gamma_{31}$ from IceCube data is stronger than the current bound by $\sim1$ order of magnitudes.

\subsection{Cascade Analysis}
\label{cascades}

Let us discuss the sensitivity of atmospheric induced cascade events to VEP parameters. As we mentioned, the cascade events originate from neutral current interaction of all neutrino flavors and charged current interaction of electron and tau neutrinos. Thus, any conversion of $\nu_\mu$ ($\bar{\nu}_\mu$) to $\nu_e$ ($\bar{\nu}_e$) or $\nu_\tau$ ($\bar{\nu}_\tau$) would lead to a distortion in the zenith and energy distributions of cascade events. Especially, as we discussed in section~\ref{sec:vep-energy}, VEP lead to $\nu_\mu\to\nu_e$ conversion and so affect the cascade distributions. Although for cascade events the zenith resolution is poor, but the better energy resolution with respect to muon-track events makes the cascade analysis plausible. 

We calculate the sensitivity of IceCube (including the DeepCore part~\cite{Collaboration:2011ym} which has higher efficiency in cascade detection) to VEP parameters. The number of cascade events can be calculated similar to section~\ref{Muon-tracks analysis} by taking into account the appropriate effective volume of IceCube for cascade detection. For the details of the calculation of cascade number of events see~\cite{Esmaili:2013cja}. Again by performing an $\chi^2$ analysis (confronting energy distribution instead of zenith distribution for muon-tracks) we estimated the sensitivity of cascades to VEP parameters. We have found a negligible increase in $\chi^2$ value compared to the standard oscillation scenario for cascades. Thus, the limit on VEP parameters from cascades are weaker than the limits from muon-tracks and the strongest limit on VEP parameter are the ones reported in section~\ref{Muon-tracks analysis}.

\section{Conclusion}\label{sec:con}

One of the essential pillars in the theory of gravitation, both classical and relativistic, is the equivalence principle which has been tested in a variety of experiments. Violation of equivalence principle has far-reaching consequences in the neutrino sector, basically introducing novel oscillation pattern which can be measured at neutrino oscillation experiments. Thus, neutrino phenomenology provides a unique tool to probe the possible violation of equivalence principle. The strength of VEP effect on neutrino oscillation depends on the neutrino energy: the VEP effectively introduce  mass-squared differences proportional to $E_\nu^2$ and so the oscillation phase will be proportional to $E_\nu$. Thus, clearly, the recent collected data of high energy ($\gtrsim100$~GeV) atmospheric neutrinos by IceCube experiment can discover/constrain VEP unprecedentedly. 

In this paper we studied the effect of VEP on the oscillation of high energy atmospheric neutrinos. In the high energy range the conventional standard oscillation induced by $\Delta m_{ij}^2$ is absent and the survival probability of each neutrino flavor is $\sim1$. However, VEP can drastically change this pattern: the effective energy-dependent mass-squared differences induced by VEP can lead to resonance flavor conversions and also oscillatory behavior in high energy with new maxima and minima in flavor oscillation probabilities. For the phenomenological model of VEP we considered in this paper, with the two VEP parameters $\phi\Delta \gamma_{21}$ and $\phi\Delta \gamma_{31}$, we studied in detail the oscillation pattern and provided the analytical descriptions of oscillation probabilities. We justified the numerical calculation of oscillation probabilities (especially $P(\nu_\mu\to\nu_\mu)$ which plays the main role in IceCube analysis) with the obtained analytical expressions and showed that the analytical approximation explains the oscillation pattern with impressive accuracy. 

Furthermore, we confronted the expected zenith distribution of muon-track events in the presence of VEP with the collected data by IceCube experiments with two different configurations, namely IC-40 and IC-79 data sets. To analyze these data we performed a simple $\chi^2$ analysis taking into account the statistical and systematic errors. The oscillation probabilities have been calculated numerically by scanning the parameter space of VEP parameters in the full three flavors framework. From these analyses we obtained the following bounds on the VEP parameters at 90\% C.L.: $-9.2\times10^{-27} < \phi\Delta\gamma_{21} < 9.1\times10^{-27}$ and $-6.3\times10^{-27} < \phi\Delta\gamma_{31} < 5.6\times10^{-27}$.  The obtained limit on $\phi\Delta\gamma_{21}$ is $\sim4$ orders of magnitude stronger than the current limit; also we improved the existing bound on $\phi\Delta\gamma_{31}$ by $\sim1$ order of magnitude.   

Finally we investigated the future sensitivity of IceCube to VEP parameters. We have presented the sensitivity region in VEP parameter space assuming three times of IC-79 data set; which improves mildly the obtained limits. Also, we have studied the effect of VEP on cascade events in IceCube, motivated by the fact that VEP induces $\nu_\mu\to\nu_e$ conversion that can distort the energy distribution of cascade events. However, due to lower statistics and higher uncertainties for cascade detection, the sensitivity of IceCube to VEP parameters in cascade channel is less than the sensitivity in muon-track channel.  
 
At the end we would like to emphasize that the limits obtained in this paper can be translated to limits on the parameters of theories (either effective theories or extensions of Standard Model) which predict/accommodate VEP to some level. As an example in this line we can mention the Standard Model Extension (SME) theories which consist of extending the Standard Model action by including all the possible terms that violate the Lorentz invariance~\cite{Colladay:1996iz,Colladay:1998fq,Kostelecky:2010ze}. One of the consequences of SME is the violation of equivalence principle such that test of VEP provide a tool for searches of Lorentz symmetry violation. Further speculations regarding these connections and possibilities to probe fundamental theories by VEP tests in neutrino sector~\cite{Kostelecky:2003cr,Kostelecky:2003xn,Kostelecky:2004hg,Diaz:2009qk,Katori:2006mz} will be pursued in a later work.

\begin{acknowledgments}
O.~L.~G.~P. thanks the ICTP for hospitality and financial support from the funding grant 2012/16389-1, S\~ao Paulo Research Foundation (FAPESP). A.~E. thanks financial support from the funding grants 2009/17924-5 from S\~ao Paulo Research Foundation (FAPESP), Jovem Pesquisador 1155/13 from FAEPEX/UNICAMP and 1280477 from PNPD/CAPES. M.~M.~G. thanks FAPESP and CNPq for several financial supports. D.~R.~G. thanks PNPD/CAPES for financial support. G.~A.~V. thanks CNPq for the financial support from the funding grant 477588/2013-1.
\end{acknowledgments}


\end{document}